\documentclass[showpacs,prb,letterpaper,twocolumn,floatfix,nobalancelastpage]{revtex4}

\usepackage{hyperref}
\hypersetup{colorlinks}
\usepackage{amsmath}
\usepackage{graphicx}
\usepackage{color}
\usepackage{calc} % math with lengths
\usepackage{xspace} % For spacing after units macros
\usepackage{todonotes}
\usepackage{overpic}

\newcommand{\nm}{\ensuremath{\mathrm{\,nm}}\xspace}

\newcommand{\mm}{\ensuremath{\mathrm{\,mm}}\xspace}

\newcommand{\GHz}{\ensuremath{\mathrm{\,GHz}}\xspace}

\newcommand{\G}[1][{}]{\ensuremath{{\overline{\mathbf G}}_\mathrm{#1}}}
\newcommand{\e}{\mathbf e}
\newcommand{\id}{\mathrm d}
\newcommand{\R}{\mathbf r}
\newcommand{\q}{\mathbf{q}}
\newcommand{\psif}{\psi_\mathrm{f}}
\newcommand{\psib}{\psi_\mathrm{b}}
\newcommand{\zerofoot}[1]{{\lefteqn{\scriptstyle{#1}}}\ }

\begin{document}
% title (fold)
\title{Theory of disorder-induced multiple coherent scattering\\ in photonic crystal waveguides}

\author{M. Patterson}
%\author{S. Combri\'e$^2$}
%\author{A. De Rossi$^2$}
\author{S. Hughes}
\email{shughes@physics.queensu.ca}
\affiliation{Department of Physics, Queen's University, Kingston, ON K7L 3N6, Canada}
%$^2$Thales Research and Technology, Route D\'epartementale 128, 91767 Palaiseau CEDEX, France}

\date{\today}

\begin{abstract}
We introduce a theoretical formalism to describe  disorder-induced extrinsic scattering in slow-light photonic crystal waveguides. This work details and  extends the optical scattering theory used in a recent \emph{Physical Review Letter} [M. Patterson \emph{et al.}, \emph{Phys.\ Rev.\ Lett.}\ \textbf{102}, 103901 (2009)] to describe coherent scattering phenomena
and successfully explain complex experimental measurements. Our presented
 theory, that combines Green function and coupled mode methods, allows one to self-consistently account for arbitrary multiple scattering for the propagating electric field and recover experimental features such as resonances near the band edge. The technique is fully
  three-dimensional and can calculate the effects of disorder on the propagating field over thousands of unit cells.
  As an application of this theory, we explore various sample lengths and disordered instances, and demonstrate the profound effect of multiple scattering
in the waveguide transmission. The spectra yield rich features associated with disorder-induced localization and multiple scattering, which are shown to be exasperated in the slow light propagation regime.
\end{abstract}

\maketitle
% title (end)

% \section introduction (fold)
Photonic crystal (PC) waveguides are structures formed by a line defect in an otherwise nominally perfect photonic crystal lattice. PC slab waveguides are of particular interest because they can be fabricated using high quality etching and lithography techniques. By guiding light using the photonic band gap of the surrounding crystal, strong transverse confinement on the order of a wavelength can be achieved. PC waveguides often exhibit a region of slow light propagation~\cite{Notomi:2001, Vlasov:2005} which has potential applications as an optical delay line \cite{Baba:2008a} or for enhanced light-matter interactions.

It is now widely accepted that slow light propagation enhances scattering from structural imperfections
or fabrication {\em disorder},  leading to significant propagation losses \cite{Kuramochi:2005,Parini:2008}. Incoherent scattering theories that calculate the loss in a single waveguide period averaged over many nominally identical samples have predicted backscattering and radiative loss to scale with the group velocity $v_g$, as $v_g^{-2}$ and $v_g^{-1}$ respectively \cite{Hughes:2005, Povinelli:2004, Gerace:2004}. These approximate loss-scaling relations have been confirmed experimentally,
 %by a number of groups,
 e.g.~\cite{Kuramochi:2005,OFaolain:2007,Engelen:2008}, but they break down at low group velocities where multiple disorder-induced scattering becomes significant. The simple scaling trends expected also typically do not include effects such as variation of the Bloch mode with wave vector or extrapolating the unit-cell loss to multiple waveguide periods, though recent work has included such effects within an incoherent scattering approach and shown a dramatic impact on the loss versus
group velocity scaling rules~\cite{Patterson:2009b}. Enhanced scattering losses
in other material systems also occur in the slow light regime, for example, massive
losses also occur in slow-light metamaterial waveguides~\cite{Arvin:Nature08}.

\begin{figure}[t]
	\centering
	\includegraphics{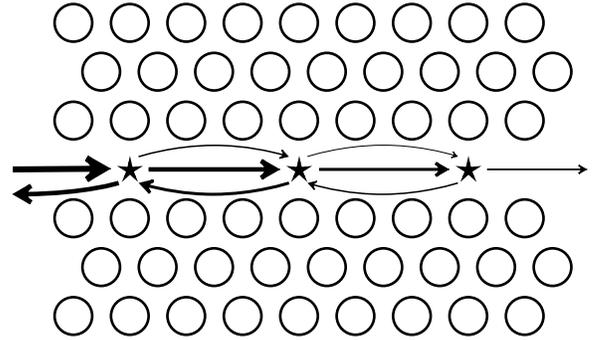}
	\caption{\label{fig:schematic} Schematic of the problem under consideration: a
plan view  of a PC waveguide with scattering sites (stars) whose strength is enhanced by the vanishing group velocity. Light injected from the left, indicated by the arrows, undergoes scattering at each of the sites leading to a complex interplay of forward and backward propagating waves. Ultimately, most of the light is back scattered and the transmission is low. In the full calculation, the scatterers are continuously distributed throughout the system and we account for the three-dimensional nature of the structure.}
\end{figure}

In a recent \emph{Physical Review Letter} \cite{Patterson:2009} by Patterson {\em et al.},
we extended previous theoretical {\em incoherent-scattering} work \cite{Hughes:2005}  to model {\em coherent scattering} over the entire length of a disordered waveguide instance, as schematically illustrated in Figure~\ref{fig:schematic}. This theory explained recent experimental reports of features such as narrow-band resonances near the band edge~\cite{Topolancik:2007,Parini:2008} and showed excellent agreement with measurements on GaAs PC structures also presented. Similar theoretical findings were later reported and confirmed by \citet{Mazoyer:2009}.
In this work, we present and expand on the theory exploited in Ref.~\citenum{Patterson:2009} and provide a full derivation.
Specifically, we introduce a non-perturbative theory of coherent optical scattering over multiple periods of a disordered waveguide instance. The theory combines Green function techniques and coupled mode formalisms with wave amplitudes calculated at each point along the length of the waveguide, where coupling coefficients include the full three-dimensional disordered structure. In Section~\ref{theory}, we introduce the theoretical formalism and derive the coupled mode equations for the forward and backward propagating  Bloch fields. In Section~\ref{sec:disorder_model}, we discuss the disorder model, and Section~\ref{sec:implementation} implements the model with examples of simulated PC waveguide transmission and forward wave intensity. Finally, we conclude in Section~\ref{sec:transmission}.
% introduction (end)

\section{Theory} % (fold)
\label{theory}
\subsection{Waveguide Bloch Modes} % (fold)
\label{sub:waveguide_bloch_modes}
The ideal PC waveguide is periodic along the propagation direction ($x$) with periodicity $a$: $\varepsilon(\R + a \hat{\mathbf x}) = \varepsilon(\R)$, where $\varepsilon(\R)$ is the dielectric constant that we will assume is real
and $\hat{\bf x}$ is a unit vector. Consequently, Bloch's Theorem applies and the electric field mode may be written as
$
	\mathbf E_k(\R) \propto \e_k(\R) \, e^{ikx},
$
where $k$ is the Bloch wave vector and $\e_k(\R)$ is the periodic Bloch mode. The magnetic Bloch mode $\mathbf h_k(\R)$ is defined similarly. Due to the Hermitian property of the Maxwell wave equations, the Bloch modes are orthogonal and, using the electric field modes, can be normalized through \cite{Sakoda:2005}
\begin{equation}
	\int_\zerofoot{\mathrm{unit\ cell}} \id \R \, \varepsilon(\R) \, \e_k^*(\R) \, e^{-ikx} \cdot \e_{k'}(\R) \, e^{ik'x} = \delta_{k,k'},
	\label{eqn:BEOrtho}
\end{equation}
where $\delta_{k,k'}$ is the Kronecker delta; a similar relation holds for the magnetic field. The use of this relation as a projection operator requires integration over the volume of a unit cell.
For the present work, since we are interested in developing sub unit-cell propagation equations, we would prefer the integration was over only the plane perpendicular to the propagation direction.
Using the electric and magnetic field orthogonality relations, the Maxwell constitutive relations, and the divergence theorem, one can derive
%can be shown that
\begin{align}
	0 &= \frac{i}{\omega \varepsilon_0} \left(1 - e^{i(k'-k)a} \right) \nonumber \\
		& \quad {} \times \iint_\zerofoot{x=x_0} \id y \, \id z \, \hat{\mathbf x} \cdot (\mathbf h_k^*(\R) \, e^{-ikx} \times \e_{k'}(\R) \, e^{ikx}),
\end{align}
where the integration here is performed over a single plane transverse to the propagation direction. For $k \ne k'$, the term in brackets is non-zero and the integral must evaluate to zero. For $k=k'$, the integral can be recognized as the power flux at the transverse plane which is clearly non-zero (except for a radiation mode propagating perpendicular to the slab). Thus, a new projection (orthogonality) operator can be defined as \cite{Patterson:MScE}
\begin{align}
	\mathcal{P}_k \mathbf E^{\rm p}(\R) &= \frac
%		{\iint_{x=x_0} \id y \, \id z \, \hat{\mathbf x} \cdot (\mathbf{h}_k^*(\R) \, e^{-ikx} \times \mathbf e^{\rm p}(\R))}
		{\iint_{x=x_0} \id y \, \id z \, \hat{\mathbf x} \cdot (\mathbf{h}_k^*(\R) \, e^{-ikx} \times \mathbf E^{\rm p}(\R))}
		{\iint_{x=x_0} \id y \, \id z \, \hat{\mathbf x} \cdot (\mathbf{h}_k^*(\R) \, e^{-ikx} \times \e_{k}(\R) \, e^{ikx})},
		\label{eqn:BProject}
\end{align}
where $\mathbf E^{\rm p}(\R)$ is the field being projected and $x=x_0$ is an arbitrary plane. This result is in agreement with that of Marcuse \cite{Marcuse:1974} and the standard form for overlap$_{\phantom A}^{}$ integrals \cite{Palamaru:2001}. The projection operator $\mathcal P_k$ has the useful property that
$
%\mathcal{P}_k {\bf E}^{\rm p} =
	\mathcal{P}_k \e_{k'}(\R) e^{ik'x} = \delta_{k,k'}.
$
% subsection waveguide_bloch_modes (end)

\subsection{Green Function Approach for the Electric Field} % (fold)
\label{sub:green_function_and_dyson_equation}
The electric-field properties of the disordered structure can be calculated analytically from Green function
solution to the electric field wave equation, namely
%the ideal properties by the Dyson equation
\begin{align}
	% Dyson equation:
	\mathbf E(\R;\omega) &= \mathbf E_i(\R;\omega) + \int_\zerofoot{\mathrm{all\ space}} \id\R' \,\G(\R,\R';\omega) \cdot \frac{\mathbf P(\R';\omega)}{\varepsilon_0},
	\label{eqn:dyson}
\end{align}
where $\mathbf P(\R';\omega)$ is the polarization density  due to the disorder in the system (defined later), $\mathbf E_i(\R;\omega)$ is the electric field in the ideal system, and $\G(\R,\R';\omega)$ is the photon Green function where the overbar represents a tensor or dyadic. The Green function is a dipole solution to the Maxwell wave equation:
\begin{equation}
	\left[ \nabla \times \nabla \times
	- \left(\frac{\omega}{c}\right)^2 \, \varepsilon(\R) \right] \G(\R,\R';\omega)
	= \left(\frac{\omega}{c}\right)^2 \, \delta(\R - \R') \, \overline{\mathbf 1},
	\label{eqn:1maxG}
\end{equation}
where $\overline{\mathbf 1}$ is the unit dyadic.
For convenience, we partition the Green function into contributions from the bound waveguide mode, radiation modes, and other modes as
\begin{equation}
	\G(\R,\R';\omega) = \G[B](\R,\R';\omega) + \G[R](\R,\R';\omega) + \G[O](\R,\R';\omega).
\end{equation}

The bound mode Green function is given analytically from properties of the bound mode \cite{Hughes:2005, Patterson:MScE}
\begin{align}
	% GBound:
	\G[B](\R,\R';\omega) &= i \frac{a \omega}{2 v_g}
		\left[  \e_{k}(\R) \otimes \e_{k}^*(\R') \, e^{ik(x-x')} \, \Theta(x-x') \right. \nonumber \\
			&\quad \left. {} + \e_{k}^*(\R) \otimes \e_{k}(\R') \, e^{ik(x'-x)} \, \Theta(x'-x) \right],
	\label{eqn:1gbound}
\end{align}
where the group velocity, $v_g$, is assumed positive (in the case of anomalous dispersion, $k$ is then negative), $\otimes$ is a tensor product, $\e_{-k}(\R) = \e_k^*(\R)$, and $\Theta(x)$ is the Heaviside step function, equal to 1 if $x>0$ and 0 is $x<0$. The mode properties can be calculated with any mode solving technique; for example, we use a freely available plane wave expansion code \cite{Johnson:2001}.

The radiation Green function, $\G[R](\R,\R;\omega)$, contains contributions from the continuum of radiation modes above the light line that are not confined to the slab by total internal reflection. The radiation Green function, whose
 contribution is significantly smaller than the dominant
 bound mode, is rather featureless and is well approximated by using a homogeneous dielectric slab with an effective permittivity determined through numerical FDTD simulations.  We
compute the radiation Green function efficiently by using the method of \citet{Paulus:2000} (see also Ref. \citenum{Patterson:MScE} for more details of our specific implementation).

The remainder of the contributions to the Green function are contained in
$\G[O](\R,\R';\omega)$ (`O' represents others), such as the possibility of having other modes (bound or leaky), and the divergence contribution of the real part of the Green function as $\R \rightarrow \R'$. Since we consider a waveguide with one bound mode in the frequency range of interest, we can safely neglect other bound modes. For the divergent contribution to $\G[O](\R,\R';\omega)$, we shall neglect its contribution in this work; the dominant effect is to cause a ridged frequency shift \cite{Ramunno:2009} and introduce local field corrections \cite{Johnson:2005, Wang:2008}.
%In what follows, for slow light PC waveguide geometries, the Green function is dominated by the bound mode contribution and the %radiation modes provide a small background effect (compare the $v_g$ pre-factors).
% subsection green_function_and_dyson_equation (end)

\subsection{Forward Wave Envelope Equation} % (fold)
\label{sub:coupled_mode_derrivation}
The electric field in the ideal waveguide can be decomposed into the complete Bloch-mode basis consisting of the target bound waveguide modes $\e_{\pm k}(\R)$, and the set of radiation modes $\{\q(\R)\}$ as
\begin{align}
	% E decomposed into modes
	\mathbf E(\R;\omega) &= \mathcal E_0 \Big[
		\e_k(\R)\,e^{ikx}\,\psi_{\rm f}(x)
		+ \e_k^*(\R)\,e^{-ikx}\,\psi_{\rm b}(x) \nonumber\\
		& \quad {} + \sum_\q \q(\R) \, e^{ik_\q x} \, \psi_\q(x) \Big],
		\label{eqn:waves}
\end{align}
where $\mathcal E_0$ is an amplitude and $\psif(x)$, $\psib(x)$, and $\{\psi_\q(x)\}$ are the envelopes for the forward, backward, and radiation modes. We stress that we use envelopes only for convenience and do not require that they are slowly varying. We are only interested in the envelopes for the bound waveguide modes but we initially track the radiation modes to include radiation scattering.

The field in a disordered waveguide can be calculated analytically from Equation \ref{eqn:dyson}, using the effective PC waveguide Green function and the disorder polarization density $\mathbf P(\R;\omega) = \varepsilon_0 \Delta\varepsilon(\R) \, \mathbf E(\R;\omega)$, as
\begin{align}
	% Dyson equation:
	\mathbf E(\R;\omega) %&=& \mathbf E_i(\R;\omega) + \int_\mathrm{all\ space} \id\R' \,\G(\R,\R';\omega) \cdot [\Delta\varepsilon(\R') \, \mathbf E(\R';\omega)] \nonumber\\
	&\simeq \mathbf E_i(\R;\omega) \nonumber \\
	&\quad {} + \int \id\R' \,\left[\,\G[B](\R,\R';\omega) + \G[R](\R,\R';\omega)\right] \nonumber \\
	&\quad {} \cdot [\Delta\varepsilon(\R') \, \mathbf E(\R';\omega)],
	\label{eqn:dyson2}
\end{align}
where $\Delta\varepsilon(\R) = \varepsilon(\R) - \varepsilon_\mathrm{i}(\R)$ is the disorder function and $\varepsilon_\mathrm{i}(\R)$ is the dielectric constant for the ideal structure. We assume an initial electric field $\mathbf E_i(\R;\omega) = \mathcal E_0 \, \e_k(\R) \, e^{ikx}$, and a total field including scattering $\mathbf E(\R;\omega)$ given by Equation~\ref{eqn:waves}.

We begin by projecting Equation~\ref{eqn:dyson2} onto a forward propagating wave by operating with $\mathcal P_k$. We then multiply by $\mathcal E_0^{-1}$ and differentiate with respect to $x$. The left hand side becomes simply $\id \psif(x) / \id x$. The projection of $\mathbf E_i(\R;\omega)$ equals $1$ and differentiating eliminates the contribution of the field in the ideal structure. This derivation will transform the integral description of the total electric field into a set of coupled propagation equations and the electric field in the ideal structure will be included as a wave injected from the input port. Equation \ref{eqn:dyson2} for the forward wave becomes
\begin{align}
	\frac{\id}{\id x} \psif(x)
	= \frac{i}{v_g} \Bigg[ &
		c_\mathrm{ff}(x) \, \psif(x) + c_\mathrm{fb}(x) \, e^{-i2kx} \, \psib(x) \nonumber \\
		& {} + \sum_\mathbf{q} c_\mathrm{f\mathbf{q}}(x) \, \psi_\mathbf{q}(x) \Bigg].
	\label{eqn:coupled_forward_partialrad}
\end{align}
The terms on the right hand side all arise from the projection of the $\G[B](\R,\R';\omega)$ term; the projection of the $\G[R](\R,\R';\omega)$ term is $0$ since the constituent radiation modes are orthogonal to the chosen bound mode. The volume integral has been converted to an integral over the transverse plane by the derivative of the Heaviside function in $\G[B](\R,\R';\omega)$. The scattering coefficients, corresponding
to {\em forward-forward}, {\em forward-backward}, and {\em forward-radiation} scatter, are
\begin{align}
	c_{\rm ff}(x) &=  \frac{a \omega}{2} \iint \id y\, \id z\, \e^*_k(\R) \cdot \e_k(\R) \, \Delta\varepsilon(\R), \label{eqn:cff}  \\
	c_{\rm fb}(x) &= \frac{a \omega}{2} \iint \id y\, \id z\, \e^*_k(\R) \cdot \e^*_k(\R) \, \Delta\varepsilon(\R), \label{eqn:cbf} \\
	c_\mathrm{f \mathbf{q}}(x) &= \frac{a \omega}{2} \iint \id y \, \id z \, \e_k^*(\R) \, e^{-ikx} \cdot \q(\R) \, e^{ik_\q x} \, \Delta\varepsilon(\R). \label{eqn:cfr}
\end{align}

An analogous equation to Equation~\ref{eqn:coupled_forward_partialrad} for $\id \psib(x) / \id x$ is formed by projecting Equation~\ref{eqn:dyson2} onto a backward propagating wave. One has
\begin{align}
	\frac{\id}{\id x} \psib(x)
	= \frac{-i}{v_g} \Bigg[ &
		c_\mathrm{bb}(x) \, \psib(x) + c_\mathrm{bf}(x) \, e^{i2kx} \, \psif(x) \nonumber \\
		& {} + \sum_\mathbf{q} c_\mathrm{b\mathbf{q}}(x) \, \psi_\mathbf{q}(x) \Bigg],
	\label{eqn:coupled_back_partialrad}
\end{align}
where the negative sign arises from the Heaviside function in Equation~\ref{eqn:1gbound}, $c_\mathrm{bb}(x) = c_\mathrm{ff}(x)$, $c_\mathrm{bf}(x) = c_\mathrm{fb}^*(x)$, and
\begin{align}
	c_\mathrm{b \mathbf{q}}(x) &= \frac{a \omega}{2} \iint \id y \, \id z \, \e_k(\R) \, e^{ikx} \cdot \q(\R) \, e^{ik_\q x} \, \Delta\varepsilon(\R). \notag
\end{align}
% subsection coupled_mode_derrivation (end)

\subsection{
%Elimination of Radiation Mode Envelopes and Derivation of  the
Disorder-Mediated Coupled Mode Equations} % (fold)
\label{sub:radiation_modes}

Next, we seek to eliminate the $\psi_q(x)$ from the equation since there are a large (infinite) number of radiation modes, and we would rather not have to solve for all the $\psi_\q(x)$.
We project Equation~\ref{eqn:dyson2} onto any one of the radiation modes to derive a radiation mode envelope equation. The left hand side becomes simply $\psi_\q(x)$. Only the $\G[R](\R,\R';\omega)$ term on the right hand side will have a non-zero projection since any chosen radiation mode will be orthogonal to the bound waveguide modes. Thus we obtain a set of equations, one for each of the radiation modes $\q$,
\begin{subequations} \label{eqn:4coupled_rad}
\begin{align}
	% Radiation coupled mode
	& \psi_\q(x) = \mathcal E_0^{-1} \mathcal P_\q
	\int \id \R' \, \G[R](\R,\R';\omega) \cdot [\mathbf E(\R';\omega) \, \Delta\varepsilon(\R')]
	\nonumber \\
	% Expand E(\R';\omega)
	&= \mathcal P_\q
		\int \id \R' \, \G[R](\R,\R';\omega) \cdot
		\e_k(\R)\,e^{ikx}\,\psi_{\rm f}(x) \, \Delta\varepsilon(\R')
		\label{eqn:4coupled_rad_f} \\
	& \quad {} \!\!+ \mathcal P_\q
		\int \id \R' \, \G[R](\R,\R';\omega) \cdot
		\e_k^*(\R)\,e^{-ikx}\,\psi_{\rm b}(x) \, \Delta\varepsilon(\R')
		\label{eqn:4coupled_rad_b} \\
	& \quad {} \!\!+ \mathcal P_\q
		\int \id \R' \, \G[R](\R,\R';\omega) \cdot
		\sum_\q \q(\R) \, e^{-ik_\q x} \, \psi_\q(x) \, \Delta\varepsilon(\R').
		\label{eqn:4coupled_rad_r}
\end{align}
\end{subequations}
There are three sources of energy for the radiation modes that are expressed as three terms on the right hand side of Equation \ref{eqn:4coupled_rad}: scattering from the forward wave (\ref{eqn:4coupled_rad_f}), scattering from the backward wave ({\ref{eqn:4coupled_rad_b}), and scattering from all the radiation modes (including self-scattering from the current radiation mode into itself) (\ref{eqn:4coupled_rad_r}).

First, we omit \ref{eqn:4coupled_rad_r}, since we assume that scattering is just a loss mechanism and inter-radiation-mode scattering will not feed back into the waveguide modes. We also neglect \ref{eqn:4coupled_rad_b}; this would give rise to radiation-assisted back-scattering where light from the backward mode scatters into a radiation mode and then the forward mode. These assumptions are reasonable because the radiation modes quickly leak from the slab and so do not interact with the scattering regions for very long. This leaves only \ref{eqn:4coupled_rad_f} which accounts for loss from the forward mode into the radiation modes. The $\mathcal P_\q$ prefix in Equation~\ref{eqn:4coupled_rad} is a projection operator acting on the radiation Green function. In Equation~\ref{eqn:coupled_forward_partialrad}, the projected Green function (in $\psi_\q$) is multiplied by the basis vector (in $c_\mathrm{f\mathbf q}$). Since the set $\{\q(\R)\}$ spans all radiation modes included in $\G[R](\R,\R';\omega)$, this is an identity transform of $\G[R](\R,\R';\omega)$ and Equation~\ref{eqn:coupled_forward_partialrad}, under substitution by Equation~\ref{eqn:4coupled_rad}, becomes
\begin{align}
	% Forward coupled mode eqn with radiation term in progress
	v_g \frac{\id}{\id x} \psif(x)
	&= i \, c_\mathrm{ff}(x) \, \psif(x)
	+ i \, c_\mathrm{fb}(x) \, e^{-i2kx} \, \psib(x) \nonumber \\
	&\quad {} + i \, c_\mathrm{fr}(x) \, \psif(x),
	\label{eqn:coupled_forward}
\end{align}
where the radiation coupling coefficient $c_\mathrm{fr}$ is given in Equation~\ref{eqn:cfr} (which is further simplified below). Note that we have conveniently eliminated the sum over~$\mathbf q$.

For the backward wave, Equation~\ref{eqn:coupled_back_partialrad} is transformed using Equation \ref{eqn:4coupled_rad} with only term \ref{eqn:4coupled_rad_b} retained. The backward wave equation is
\begin{align}
	-v_g \frac{\id\psi_{\rm b}(x)}{\id x}
	&= i\, c_{\rm bb}(x)\, \psi_{\rm b}(x)
		+ i\, c_{\rm bf}(x)\, e^{i2kx}\, \psi_{\rm f}(x) \nonumber \\
		& \quad {} + i\, c_{\rm br}(x)\, \psi_{\rm b}(x). \label{eqn:coupled_back}
\end{align}

% subsection radiation_modes (end)

%\subsection{Disorder-Induced Coupled Mode Equations} % (fold)
%\label{sub:final_equations}

The final coupled mode equations are Equations \ref{eqn:coupled_forward} and \ref{eqn:coupled_back}. The coupling coefficients can be physically interpreted as $c_{\rm ff} = c_{\rm bb}$ (\ref{eqn:cff})  driving scattering from a mode into itself,
$c_\mathrm{bf} = c_\mathrm{fb}^*$ (\ref{eqn:cbf}) driving scattering into the counter-propagating mode, and $c_{\rm fr}$ and $c_{\rm br}$  driving scattering from the waveguide mode into radiation modes above the light line.
With the elimination of the radiation mode
envelopes, the coupling coefficients into radiation modes (e.g., \ref{eqn:cfr}) become
\begin{align}
	%c_{\rm ff}(x) &= \frac{a \omega}{2} \iint \id y\, \id z\, \e^*_k(\R) \cdot \e_k(\R) \, \Delta\varepsilon(\R),
%		\label{eqn:cff} \\
	%c_\mathrm{bb}(x) &=& c_\mathrm{ff}(x),
	%	\label{eqn:cbb} \\
%	c_{\rm bf}(x) &= \frac{a \omega}{2} \iint \id y\, \id z\, \e_k(\R) \cdot \e_k(\R) \, \Delta\varepsilon(\R),
%		\label{eqn:cbf} \\
	%c_{\rm fb}(x) &=& c_\mathrm{bf}^*(x),
	%	\label{eqn:cfb} \\
	c_\mathrm{fr}(x) &=
		\frac{a \omega}{2} \iint \id y \, \id z \, \int_\mathrm{all\ space} \!\!\!\!\!\!\!\!\!\!\!\!\!\!\! \id \R' \,
		\Delta\varepsilon(\R)  \, \Delta\varepsilon(\R') \nonumber \\
		&\quad {} \times e^{-ikx} \, \e_k^*(\R)
		\cdot \G[R](\R,\R';\omega)
		\cdot \e_k(\R') \, e^{ikx'},
		\label{eqn:cfr2} \\
	c_\mathrm{br}(x) &=
		\frac{a \omega}{2} \iint \id y \, \id z \, \int_\mathrm{all\ space} \!\!\!\!\!\!\!\!\!\!\!\!\!\!\! \id \R' \,
		\Delta\varepsilon(\R)  \, \Delta\varepsilon(\R') \nonumber \\
		& \quad {} \times e^{ikx} \, \e_k(\R)
		\cdot \G[R](\R,\R';\omega)
		\cdot \e^*_k(\R') \, e^{-ikx'}.
		\label{eqn:cbr2}
\end{align}
Importantly, this theory incorporates the full three-dimensional structure of the waveguide, Bloch modes, and disorder functions in calculating the scattering.

The radiation scattering coefficients of Equations~\ref{eqn:cfr2}--\ref{eqn:cbr2} are difficult to evaluate due to the integral over the entire waveguide. Although we assume disorder between holes is uncorrelated in the expectation sense, for any instance of disorder, there may be a non-zero correlation between holes mediated by radiation modes.
 However, we are primarily interested in coherent scattering that is contained
 within the waveguide, and can reasonably assume that any field scattered
 out of a bound mode will not be scattered back into a bound mode; this is justified
  as the bound mode scattering channel is by far the dominant one. Therefore,  we
   can simply the radiation loss by using $c_\mathrm{fr} = i \, \langle \alpha_\mathrm{rad} \rangle \, v_g / 2 \, a$ where $\langle \alpha_\mathrm{rad} \rangle$ is the incoherent average radiation loss \cite{Hughes:2005}
\begin{align}
	\langle \alpha_\mathrm{rad} \rangle =&
	\frac{a \omega}{v_g} \iint \id \R' \, \id \R'' \, \langle \Delta\varepsilon(\R') \, \Delta\varepsilon(\R'') \rangle \, \e_k^*(\R') \, e^{-ikx'}\nonumber \\
		& {}  \cdot \mathrm{Im} \left[ \G[rad](\R',\R'';\omega) \right] \cdot \e_k(\R'') \, e^{ikx''}.
	\label{eqn:3alphaRadAvg}
\end{align}
Comparing Equations~\ref{eqn:3alphaRadAvg} and~\ref{eqn:cfr}, the former is just the expectation value of the imaginary part the later integrated over a unit cell. The factor of 2 is necessary to convert from a power loss to an amplitude loss.
% subsection final_equations (end)

For modelling an incident field at one end of the waveguide, the boundary conditions for a wave injected into the waveguide (and consistent with $\mathbf E_i(\R;\omega)$) are
\begin{eqnarray}
	\psif(x_\mathrm{start}) &=& 1, \\
	\psib(x_\mathrm{end}) &=& 0,
\end{eqnarray}
where $x_\mathrm{start}$ and $x_\mathrm{end}$ are the positions of the input and output ports.
The propagating envelopes are then computed at all spatial position within the waveguide
using the presented coupled mode equations (Eqs.~\ref{eqn:coupled_forward}-\ref{eqn:coupled_back}).
We stress that the full three-dimensional Bloch mode and disordered holes are
self-consistently included in these final coupled-mode equations.

% section theory (end)

\section{Disorder Model} % (fold)
\label{sec:disorder_model}
\begin{figure}
	\centering
	% Use grid option to overpic go get the coordinates for the labels
	\begin{overpic}[width=2.9in]{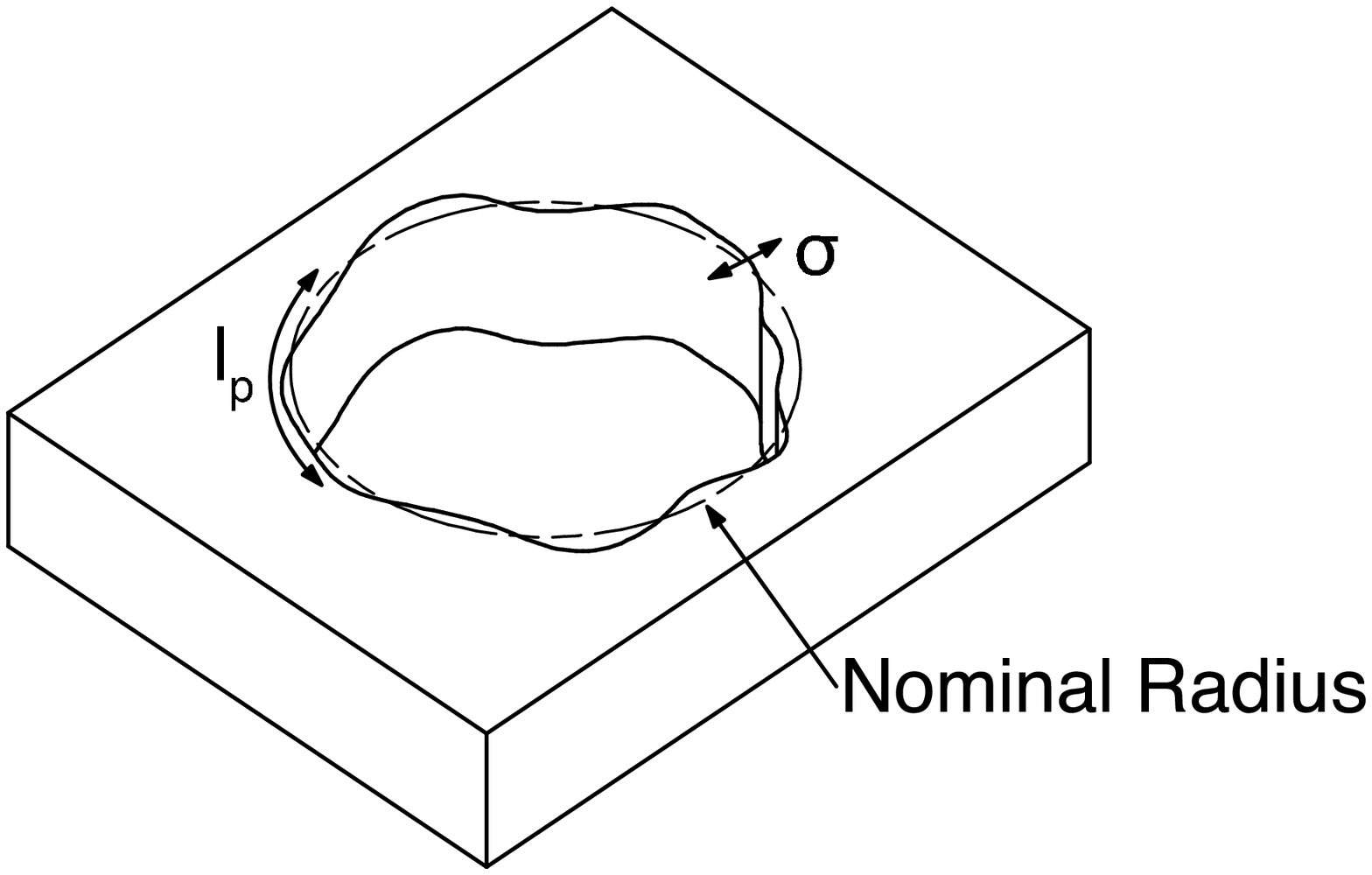}\put(10,71){a)}\end{overpic}
%	\quad
\vspace{.5cm}\\
	\begin{overpic}[width=1.9in]{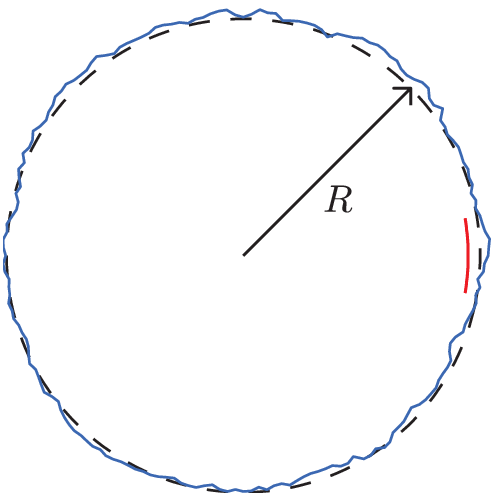}\put(-12,95){b)}\end{overpic}
	\caption[Schematic of a disordered hole]{\label{fig:RoughHole} a) Schematic of a hole with disordered perimeter and straight side walls. We describe the statistical properties of the disorder with a RMS roughness $\sigma$ and a correlation length $l_p$ measured around the circumference. b) Example of  a disordered hole profile used in the calculation (blue). The ideal radius (dashed black) and correlation length (short red arc) are shown for reference,
and $R$ indicates the nominal radius of the unperturbed hole.}
\end{figure}

%\subsection{Ideal Structure} % (fold)
%\begin{figure}
%	\centering
%	% Use grid option to overpic go get the coordinates for the labels
%	\begin{overpic}{dispersion} \put(0,65){a)}\end{overpic}
%	\vspace{.5cm}\\
%	\begin{overpic}{mode} \put(-1,20){b)}\end{overpic}
%	\caption{\label{fig:bandstructure} Properties of the nominal structure. a) Dispersion of the waveguide mode (blue, solid, left scale). The continuum of radiation mode above the light line is indicated by the shading on the left side of the figure. The group index (green, dashed, right scale) is also shown and diverges at the band edge ($k=2\pi/a$). b) Distribution of the transverse component of the electric field Bloch mode at the middle of the slab near the band edge.}
%\end{figure}

%
The equations can now be used with any disorder model. In our experience~\cite{Kuramochi:2005,Patterson:2009b} and in agreement with the analysis of images of PC slabs~\cite{Skorobogatiy:2005}, we have found that disorder in PC slab structures is dominated by perturbations of the perimeter of the holes, as shown in Figure~\ref{fig:RoughHole}. We take the radial perturbation $\Delta r$ to be a Gaussian random variable with a mean of 0 and a standard deviation of $\sigma$. Two radial perturbations are correlated by
\begin{equation}
	\langle \Delta r_i(\phi_i) \Delta r_j(\phi_j') \rangle = \sigma^2 \, e^{-R|\phi_i - \phi_j'|/l_p} \, \delta_{i,j},
	\label{eqn:disorder}
\end{equation}
where the subscript indexes the holes, $\phi_i$ is the angular position of the point measured about the centre of the hole, $R$ is the ideal hole radius, and $l_p$ is the correlation length measured around the circumference.

The change in dielectric constant about a single hole $i$ is given exactly by
\begin{align}
	\Delta\varepsilon_i(r_i,\phi_i) = {}& (\varepsilon_2 - \varepsilon_1) \, \left [ \Theta(r_i-R) \right . \nonumber \\
	& {} - \left . \Theta\left (r_i - R - \Delta r_i(\phi_i) \right) \right ],
	\label{eqn:disorder_step}
\end{align}
where $(r_i, \phi_i)$ are cylindrical coordinates centred about hole $i$. This form holds for both positive and negative values of $\Delta r_i(\phi_i)$.
The disorder $\Delta\varepsilon$ appears in the formalism in spatial integrals where it is multiplied by functions of the electric fields and Green function. We consider such an integration, where $f(r_i,\phi_i)$ represents one of the fields and is slowly varying over the relevant length scale. The field $f(r_i,\phi_i)$ can be expanded in a Taylor series along the radial coordinate to evaluate the integral as
\begin{align}
	\int & \id r_i \, \Delta \varepsilon_i(r_i, \phi_i) \, f(r_i, \phi_i) \notag \\
	& {} = \int \id r_i \, \Delta \varepsilon(r_i, \phi_i)  \notag \\
		& \quad {} \times (f(R, \phi_i) + f'(R, \phi_i) (r_i - R)
		+ O((r_i-R)^2)) \notag \\
	& {} = f(R, \phi_i) \, \int \id r_i \, \Delta \varepsilon(r_i, \phi_i) \notag \\
		& \quad + f'(R, \phi_i) \int \id r_i \, \Delta \varepsilon(r_i, \phi_i) (r_i - R)
		+ O((r_i-R)^2) \notag \\
	& {} = f(R, \phi_i) \, (\varepsilon_2 - \varepsilon_1) \, \Delta r_i(\phi_i) \notag \\
		& \quad {} + f'(R, \phi_i) \, (\varepsilon_2 - \varepsilon_1) \, \frac{\Delta r_i(\phi_i)^2}{2}
		+ O(\Delta r_i(\phi_i)^3).
	\label{eqn:depsilon_expand_temp}
\end{align}
To include the disorder to first order in $\Delta r_i(\phi_i)$, it is sufficient to take the field at the ideal hole radius $f(R, \phi_i)$. For convenience of notation, we then rewrite Equation~(\ref{eqn:disorder_step}) as
\begin{align}
	\Delta\varepsilon_i(r_i,\phi_i) = (\varepsilon_2 - \varepsilon_1) \, \delta(r_i-R) \, \Delta r_i(\phi_i),
\end{align}
so that
\begin{align}
	\int & \id r_i \, \Delta \varepsilon_i(r_i, \phi_i) \, f(r_i, \phi_i) = f(R, \phi_i) \, (\varepsilon_2 - \varepsilon_1) \, \Delta r_i(\phi_i),
	\notag
\end{align}
which agrees with Equation~(\ref{eqn:depsilon_expand_temp}) to first order.

% section disorder_model (end)

\section{Implementation} % (fold)
\label{sec:implementation}

\subsection{Ideal Structure} % (fold)
\begin{figure}
	\centering
	% Use grid option to overpic go get the coordinates for the labels
	\begin{overpic}{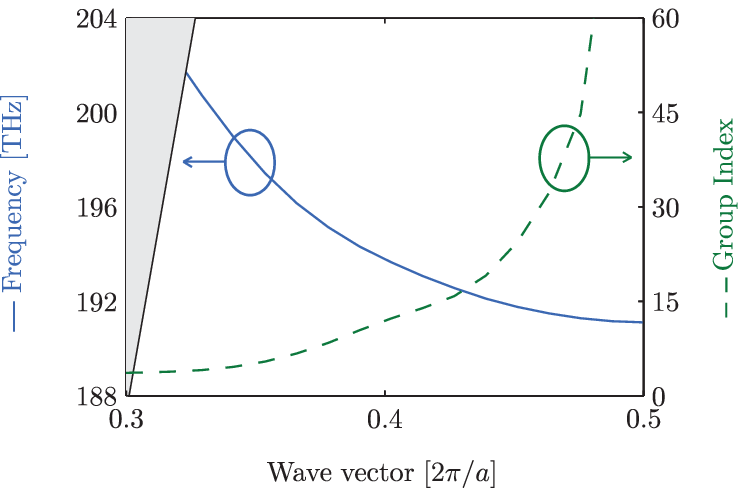} \put(0,65){a)}\end{overpic}
	\vspace{.5cm}\\
	\begin{overpic}{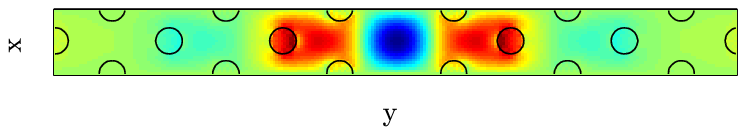} \put(-1,20){b)}\end{overpic}
	\caption{\label{fig:bandstructure} Properties of the nominal structure. a) Dispersion of the waveguide mode (blue, solid, left scale). The continuum of radiation mode above the light line is indicated by the shading on the left side of the figure. The group index (green, dashed, right scale) is also shown and diverges at the band edge ($k=2\pi/a$). b) Distribution of the transverse component of the electric field Bloch mode at the middle of the slab near the band edge.}
\end{figure}

This calculation requires, as inputs, the ideal waveguide mode dispersion and spatial field distribution. As a
 representative example we consider a W1 semiconductor waveguide with pitch $a=480\nm$, slab thickness $h=160\nm$, hole radius $R = 95\nm$, and index of refraction $n=3.18$. The dispersion of the waveguide mode is shown in Figure \ref{fig:bandstructure}a) (blue, solid, left scale) along with the group index (green, dashed, right scale). Near the band edge ($k=2\pi/a$), the group index is large, increasing scattering as the light slows down. The spatial distribution of the electric field in the centre of the slab is shown in Figure \ref{fig:bandstructure}b).}
% subsection Ideal Structure (end)

\subsection{Numerical Implementation} % (fold)
To solve Equations \ref{eqn:coupled_forward} and \ref{eqn:coupled_back} numerically, the coupling coefficients are assumed to be constant over a short ($\Delta x \ll a$) interval in $x$ and are integrated analytically. This yields a pair of transfer equations linking the envelopes on either side of the chosen interval. In this way, a set of transfer equations that span the entire waveguide length can be built, and then solved using linear algebra techniques. This approach is particularly amenable to adding reflective facets and other features by simply including an appropriate transfer matrix.

\begin{figure}
	\centering
	\includegraphics{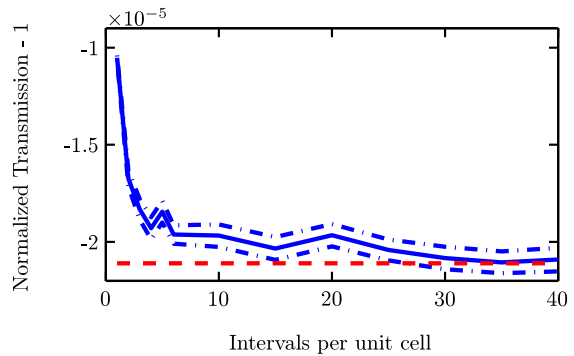}
	\caption{\label{fig:numXCell} Mean transmission through 500 disordered waveguides (blue, solid) as a function of the number of intervals each unit cell is divided into. The error in the mean is marked by the dash-dotted limits and the mean agrees well with the prediction of the incoherent calculation \cite{Hughes:2005} (red, dashed), except for very coarse discretizations.}
\end{figure}

The average coupling constants for each interval are calculated by, for each hole, generating an instance of a disordered profile from the statistical distribution of Equation \ref{eqn:disorder}. The coupling coefficients are calculated at multiple points within the interval, and then averaged. Typically, there are 20 intervals per unit cell to satisfy the assumption that the coefficients are relatively constant. As shown in Figure \ref{fig:numXCell}, if the discretization of the unit cell is too coarse, the loss is underestimated. Thus, one must include sub unit-cell propagation effects.

We highlight that
the calculation is orders of magnitude more efficient than standard brute-force numerical techniques, e.g., FDTD.
 We also note that we only need to calculate the coupled mode coefficients wherever
disorder has an influence, namely at the hole interfaces. However, the
 final computation, though efficient, is not instantaneous. Producing a high resolution transmission spectrum (1000 frequency points) for a 1\mm waveguide (2\,500 unit cells and 50\,000 grid points) takes approximately 1 cpu day (on a 2.4\GHz AMD Opteron processor).  However the calculations at each frequency are independent and the total calculation can also be greatly accelerated by exploiting parallelism. In contrast, we estimate that a minimum of about 40\,GBytes of memory and 5800 cpu days are required to perform the simulation using FDTD. Clearly, this semi-analytic treatment is a significant advantage.

% FDTD calculation:
% 20nm grid size. Simulation 1mm x 8 sqrt(3) 400nm x 1um = 962e6 gridpoints.
% Simulation time: Length 1mm / c * 100 = 333564 fs, dt = 0.0381fs => timesteps = 8.74e6.
% Total number of nodes to calculate = 6e15. Solver speed = 12M nodes / s => time = 5800 days.
% subsection numerical implementation (end)

% section implementation (end)

\section{Computed transmission spectra} % (fold)
\label{sec:transmission}
\begin{figure}
	\centering
%	 \includegraphics[width=3.3in]{coherentScatterSimulations}
%	\begin{overpic}{coherentScatterSimulations}
%		\put(12.25,94.25){a)} \put(57.75,94.25){b)}
%		\put(12.25,71){c)}    \put(57.75,71){d)}
%		\put(12.25,47.75){e)} \put(57.75,47.75){f)}
%		\put(12.25,24.5){g)}  \put(57.75,24.5){h)}
%	\end{overpic}
%
\begin{overpic}[width=3.2in]{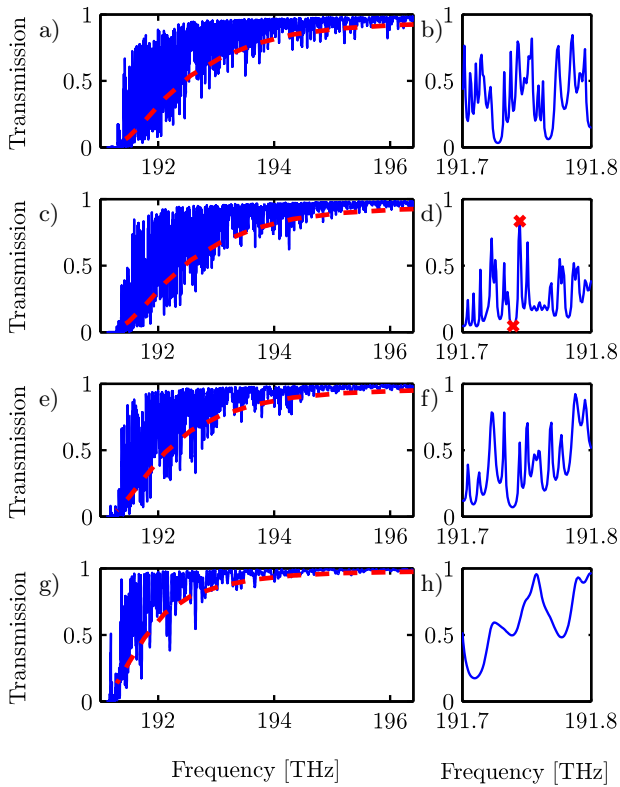}
		\put(4,94.25){a)} \put(52,94.25){b)}
		\put(4,71){c)}    \put(52,71){d)}
		\put(4,47.75){e)} \put(52,47.75){f)}
		\put(4,24.5){g)}  \put(52,24.5){h)}
		% was 12.25, ?				57.75, ?
	\end{overpic}
	\caption{\label{fig:coherentScatterSimulations} Simulated transmission spectra of four disordered W1 waveguides using the new coherent scattering theory (blue, solid) and the first- and second-order Born incoherent theory (red, dashed) \cite{Hughes:2005}. Each row of plots is for a different waveguide with the left plot showing a broad frequency range and the right plot showing a narrow frequency range near the band edge. Plots (a) and (b) are for a disordered 1.5\mm waveguide. Plots (c) and (d) are for a different disorder instance of the same 1.5\mm waveguide. Plots (e-f) and Plots (g-h) are for the same disorder instance as (c-d) but with the length reduced to 1.0\mm and 0.5\mm respectively. The calculation uses a RMS roughness of $\sigma = 3\nm$, and a disorder correlation length of $l_p = 40\nm$. The forward wave intensity as a function of position is given in Figure~\ref{fig:intensityPosition} for the two points marked with crosses in d).}
\end{figure}
Figure~\ref{fig:coherentScatterSimulations} shows transmission spectra for four disordered waveguides calculated by solving Equations~\ref{eqn:coupled_forward} and~\ref{eqn:coupled_back} (blue, solid). For reference, previous incoherent scattering results, computed within a second-order Born approximation~\cite{Hughes:2005}, are  also shown (red, dashed); we also note that extensions to the incoherent scattering theory to account for multiple scattering have been introduced recently~\cite{Patterson:2009b}. Each row of plots is for a different waveguide with the left plot showing a broad frequency range and the right plot showing a narrow frequency range near the band edge. The top row is for a disordered 1.5\mm waveguide (3125 unit cells). The second row is for a different disorder instance of the same 1.5\mm structure. Experimentally, this would be similar to carrying out measurements on a second waveguides fabricated with nominally identical parameters. It has the same qualitative shape but the particular disordered resonances are substantially different. This is important if it was desired to take advantage of these sharp resonances since their resonant frequency cannot be easily designed. The third and forth rows are for the same disorder instance as the second but with the length reduced to 1.0\mm and 0.5\mm respectively. Here the qualitative roll off changes due to the length reduction but disordered resonances can be found at similar frequencies across the three lengths, especially between the 1.5\mm and 1.0\mm cases.

% There is a pronounced change in the roll-off as a function of length; short waveguides have a sudden increase in loss near the band edge while for long waveguides, the roll-off starts farther from the band edge but is more gradual. This is expected since the transmission scales as $e^{-\alpha L}$ and so doubling the length is roughly equivalent to squaring the transmission.

\begin{figure}
	\centering
	\includegraphics[width=\columnwidth]{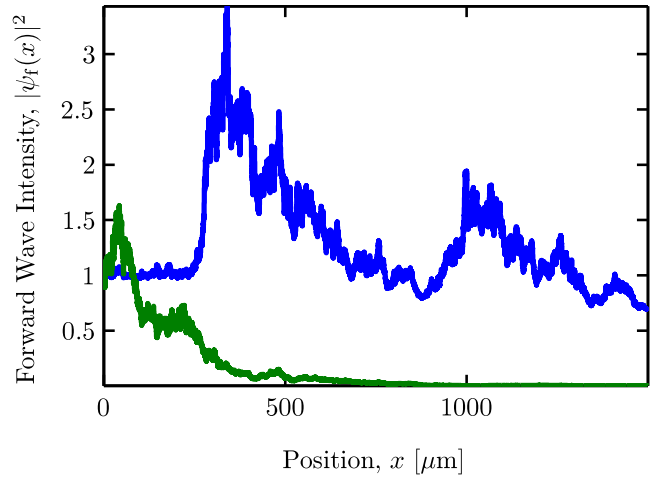}
	\caption[Wave intensity vs.\ position in disordered waveguide]{\label{fig:intensityPosition} Forward wave intensity in a disordered waveguide at two wave vectors. The blue curve ($n_g=24.96$) corresponds to a local transmission maximum and the green curve ($n_g=25.11$) is a neighbouring transmission minimum. These two curves correspond to the red crosses in the Figure~\ref{fig:coherentScatterSimulations}d).}
\end{figure}

We can examine the position-dependent distribution of energy in the waveguide under c.w.\ illumination. In the second row, right column of Figure~\ref{fig:coherentScatterSimulations}, a neighbouring transmission minimum and maximum are marked with red crosses. The forward wave intensity at these frequencies is plotted in Figure~\ref{fig:intensityPosition}. Although the points are very close in frequency, the minute difference in group index ($n_g=25.11$ compared to $n_g=24.96$) creates a difference in the accumulated phase and a dramatic change in the transmission.

By including multiple, coherent scattering we reproduce the experimental phenomenon of sharp spectral resonances near the band edge. Although initially unexpected, these features are just Fabry-P\'erot-like fringes between extrinsic scattering sites. The slow group velocity enhances scattering to create the scattering sites and also increases the effective cavity length between sites, narrowing the resonance line-width.\\
% section transmission spectra (end)

\section{Conclusions} % (fold)
\label{sec:conclusions}
We have described and applied a theory for self-consistently modelling coherent scattering in a disordered PC waveguide instance,
allowing one to map directly onto a realistic experimental situation. Slow light propagation enhances back scattering (and, to a lesser extend, radiation scattering) leading to high losses near the band edge. The formation of sharp spectral resonances near the band edge is shown which is mediated by Fabry-P\'erot-like resonances between disorder sites. This theory is computationally efficient, making the analysis of very long waveguides (thousands of periods
using the full three-dimensional structure) feasible on a desktop computer. Although the presented model may not be quantitatively exact (e.g., it neglects local field effects), the qualitative results such as the formation of sharp resonances near the band edge certainly can, and already have been, used to
explain  a rich range of experimental features without introducing any fitting parameters \cite{Patterson:2009}. The role of local field effects will be reported in future work, and the effects on incoherent frequency shifts are described elsewhere \cite{Patterson:FreqShift}.
% (end)

\section*{Acknowledgments}
This work was supported by the National Sciences and Engineering Research Council of Canada,
and the Canadian Foundation for Innovation. We thanks S. Combri{\'e}
and A. De Rossi for many useful discussions.

% Choose one of the below bibstyles or comment both to use the revtex4 default
% \bibliographystyle{plainnat} % To display all bibliographic information
% \bibliographystyle{myapsrev} % To display highly condensed bibliography (and italicised et al)
% \bibliography{../../Masters/JournalsShort,../../Masters/Masters,./coherentPaper}

\end{document}